\documentclass[a4paper,11pt]{article}
\usepackage{jinstpub} % for details on the use of the package, please
\usepackage{subfigure}

\title{Design and Commissioning of the PandaX-4T Cryogenic Distillation System for Krypton and Radon Removal}

\author[a,c]{Xiangyi Cui,}
\author[b,c,1]{Zhou Wang,\note{Corresponding author.}}
\author[d,1]{Yonglin Ju}
\author[d]{Xiuli Wang,}
\author[d]{Huaxuan Liu,}
\author[b]{Wenbo Ma,}
\author[a,b,c]{Jianglai Liu,}
\author[b,c]{Li Zhao,}
\author[a,b,e]{Xiangdong Ji}
\author[a]{Shuaijie Li,}
\author[d]{Rui Yan,}
\author[d]{Haidong Sha,}
\author[d]{Peiyao Huang,}
\author[f]{Jifang Zhou,}
\author[f]{Changsong Shang,}
\author[f]{Liqiang Liu}

\affiliation[a]{Tsung-Dao Lee Institute, Shanghai Jiao Tong University, \\Shanghai 200240, China}
\affiliation[b]{ INPAC and School of Physics and Astronomy, Shanghai Jiao Tong University, Shanghai Laboratory for Particle Physics and Cosmology,\\Shanghai 200240, China}
\affiliation[c]{Shanghai Jiao Tong University Sichuan Research Institute,\\Chengdu 610000, China}
\affiliation[d]{Institute of Refrigeration and Cryogenics, Shanghai Jiao Tong University\\Shanghai 200240, China}
\affiliation[e]{ Department of Physics, University of Maryland, \\College Park, Maryland 20742, USA}
\affiliation[f]{ Yalong River Hydropower Development Company, Ltd., \\Chengdu 610051, China}

\emailAdd{wangzhou0303@sjtu.edu.cn}
\emailAdd{yju@sjtu.edu.cn}

\abstract{An online cryogenic distillation system for the removal of krypton and radon from xenon was designed and constructed for PandaX-4T, a highly sensitive dark matter detection experiment. The krypton content in a commercial xenon product is expected to be reduced by 7 orders of magnitude with 99\% xenon collection efficiency at a flow rate of 10~kg/h by design.
The same system can reduce radon content in xenon by reversed operation, with an expected radon reduction factor of about 1.8 in PandaX-4T under a flow rate of 56.5~kg/h. The commissioning of this system was completed, with krypton and radon operations tested under respective working conditions. The krypton concentration of the product xenon was measured with an upper limit of 8.0~ppt.
}

\keywords{Distillation; Kr removal; Rn removal; Dark Matter; PandaX}

\begin{document}
\maketitle
\flushbottom

\section{Introduction}
\label{sec:intro}

In recent years, liquid xenon has been a leading target in the dark matter direct detection experiments~\cite{xe10,xe100,XMass1,zeplin,lux,panda1,panda2_final,xe1t}. The PandaX project consists of a series of xenon-based experiments located at the China Jinping Underground Laboratory (CJPL)~\cite{CJPL,CJPL-II}. The PandaX-I and PandaX-II are the two completed experiments, utilizing gas-liquid dual-phase xenon Time Projection Chambers (TPCs) with targets of 120~kg and 580~kg, respectively. PandaX-4T is the new generation experiment, which has a target of 4-ton liquid xenon. The expected sensitivity of PandaX-4T is better than $10^{-47}$~cm$^2$ to dark matter-nucleon spin-independent cross section for a dark matter mass around 40~GeV/c$^2$~\cite{panda4_need}, a 10-fold improvement in comparison to PandaX-II.

$^{85}$Kr and $^{222}$Rn are the two most important intrinsic contaminants of radioisotopes in liquid xenon detectors. $^{85}$Kr, continually increased in the atmosphere~\cite{kr_increase}, has a half-life of 10.76 years. At present time, its concentration in natural krypton is about $ 2\times10^{-11}$ mol/mol. The different physical properties between xenon and krypton make it possible to separate krypton from xenon, via distillation~\cite{kr83m} or adsorption~\cite{xe10_chro}.  The cryogenic distillation method, based on the different boiling points of krypton and xenon, was used in XMASS~\cite{XMass_kr1}, PandaX-II~\cite{panda2_dis1,panda2_dis2}, XENON100~\cite{xenon100_dis}, and has achieved a reduction of krypton level by three orders of magnitude for one pass. The upgraded distillation system of XMASS could reduce the amount of krypton by five orders of magnitude with one pass~\cite{XMass_kr2}. In the recent result of the XENON1T distillation system~\cite{xenon1t_dis}, the krypton reduction factor of 6.5$\times$10$^{5}$ is achieved. On the other hand, the LUX experiment used the adsorption-based gas chromatography method, which decreased the krypton concentration by a Kr reduction factor of 3$\times$10$^{4}$ for each batch~\cite{LUX_radio,LUX_kr}.

$^{222}$Rn is a decay progeny of $^{238}$U, therefore is emanating continuously from detector materials, pipes, and parts in direct contact with xenon. Radon can easily diffuse into the liquid xenon, then distributes homogeneously in the target volume as its half-life is 3.8 days. One of the daughter isotopes, $^{214}$Pb, is a $\beta$-emitter and an important background in the liquid xenon detector. The achieved $^{222}$Rn activity is 10.3~$\mu$Bq/kg in XMASS~\cite{XMass1}, 4.5~$\mu$Bq/kg in XENON1T experiment~\cite{xenon1t_newRn}, and 32~$\mu$Bq/kg ($^{214}$Pb) in LUX~\cite{LUX_radio}.
The XENON100 collaboration demonstrated the feasibility of radon removal by online distillation at a flow rate of 4.5 standard liters per minute (slpm), under which a radon reduction factor of 27 was achieved~\cite{xenon_rn}. On the other hand, XMASS and LZ have developed adsorption-based gas chromatography method~\cite{xmass,lz} to reduce the radon concentration.

In order to improve the sensitivity to search for dark matter by more than one order of magnitude, the PandaX-4T experiment requires a krypton concentration of less than 0.1~ppt~\cite{panda4_need}. This specification poses significant technical challenge to the distillation tower. The $^{222}$Rn concentration in PandaX-II, on the other hand, was measured to be $\sim$32~$\mu$Bq/kg (3.3$\times$10$^{-24}$~mol/mol) ~\cite{panda2_final,laoma}. Additional reduction method in radon level is desired as well.

This paper presents a high efficiency cryogenic distillation column which has been initially designed and constructed for the krypton removal, with an increased purification speed of 10~kg/h and an  expected reduction factor of $10^{7}$ for krypton. Reversed radon distillation, compared to the krypton removal, with the same column is considered in design. The rest of this paper is organized as follows. The design of the distillation column is described in section~\ref{sec:2}. The hardware setup and standard operation modes are discussed in section~\ref{sec:3} and section~\ref{sec:4}. The commissioning of the column and results are discussed in section~\ref{sec:5}, before we conclude in section~\ref{sec:6}.

\section{Design of the column}
\label{sec:2}

The boiling points of the krypton, radon and xenon are 120 K, 211 K and 165 K under the atmosphere pressure, respectively. This indicates that the krypton and radon removal process could be implemented at the same temperature near the xenon boiling point. During the krypton distillation operation, krypton is the more volatile component, which is enriched in the gaseous phase. Conversely, xenon is the more volatile component compared to radon in the radon removal process, so that radon will be enriched in the liquid phase.

\subsection{Distillation method for the Kr and Rn}
\label{sec:2:1}

\begin{figure}[htbp]
\centering 
\includegraphics[width=.7\textwidth]{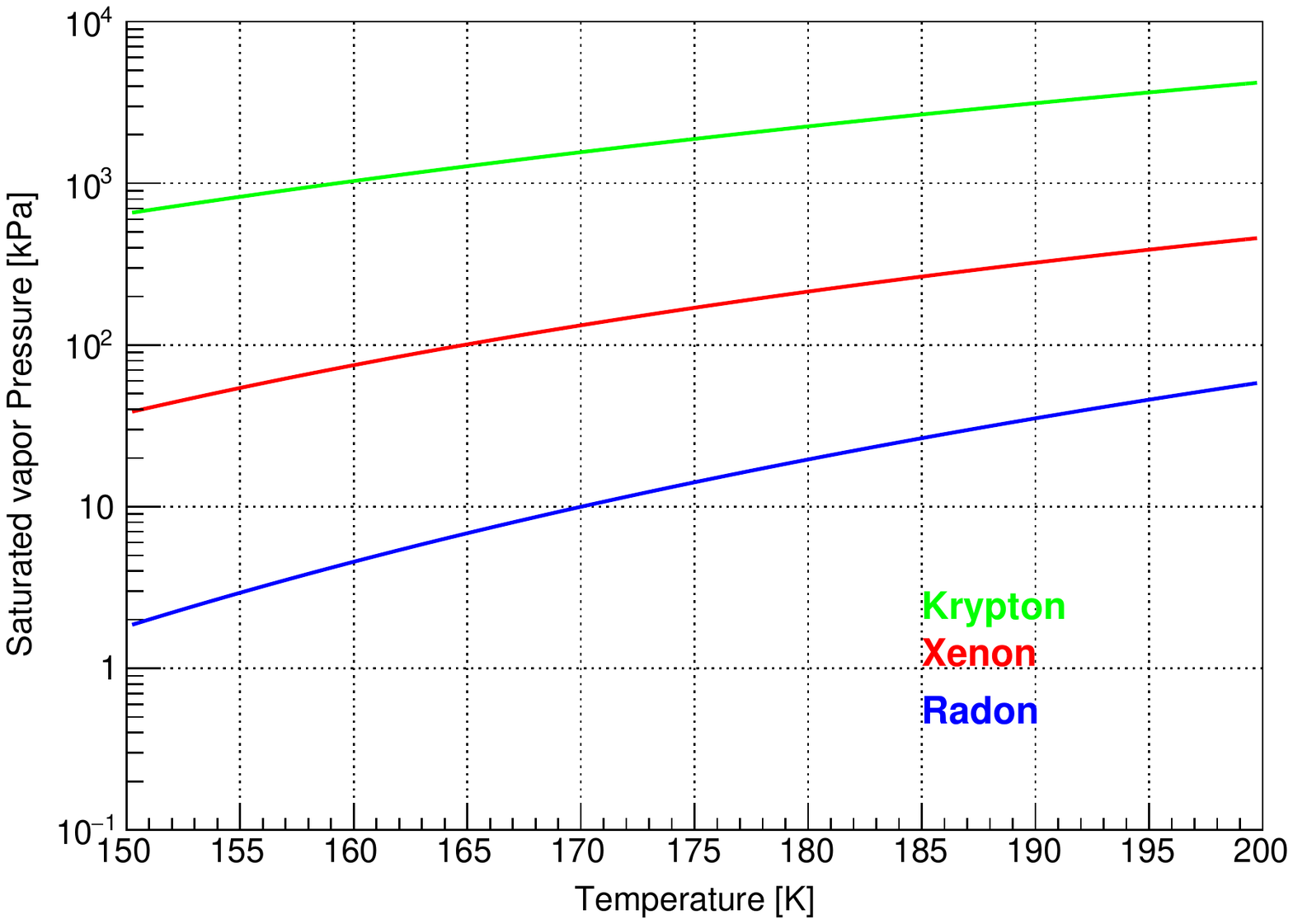}
\caption{\label{fig:2:1:1} Saturation vapor pressure curves for Kr, Rn, Xe.}
\end{figure}

The separation power to different gas species by the distillation can be quantified by the relative volatility $\alpha$, which is the ratio of the saturation vapor pressures. The saturation vapor pressure curves versus temperature for krypton, xenon and radon  are shown in figure~\ref{fig:2:1:1}~\cite{textbook,NIST}. The operation temperature is chosen to be 178~K, considering the narrow temperature range of the liquid xenon phase. At this temperature, the ratio of volatility between Kr and Xe ($\alpha^{\rm Kr})$ is 10.7
%10.4
and that between Xe and Rn ($\alpha^{\rm Rn})$ is 11.3.
%13.8. 

Raoult's law and Dalton's law are used for the ideal gas and liquid to derive the concentration of the volatile component at equilibrium state. The relationship of the volatile component concentration in the gaseous phase ($y$) and in the liquid phase ($x$) at the state of equilibrium is defined by eq.~\eqref{eq:2:1:1}:

\begin{equation}
\label{eq:2:1:1}
\begin{split}
y &= \frac{\alpha \cdot x}{1+ \left( \alpha -1 \right) \cdot x} \,.
\end{split}
\end{equation}

\begin{figure}[htbp]
\centering 
\includegraphics[width=.55\textwidth]{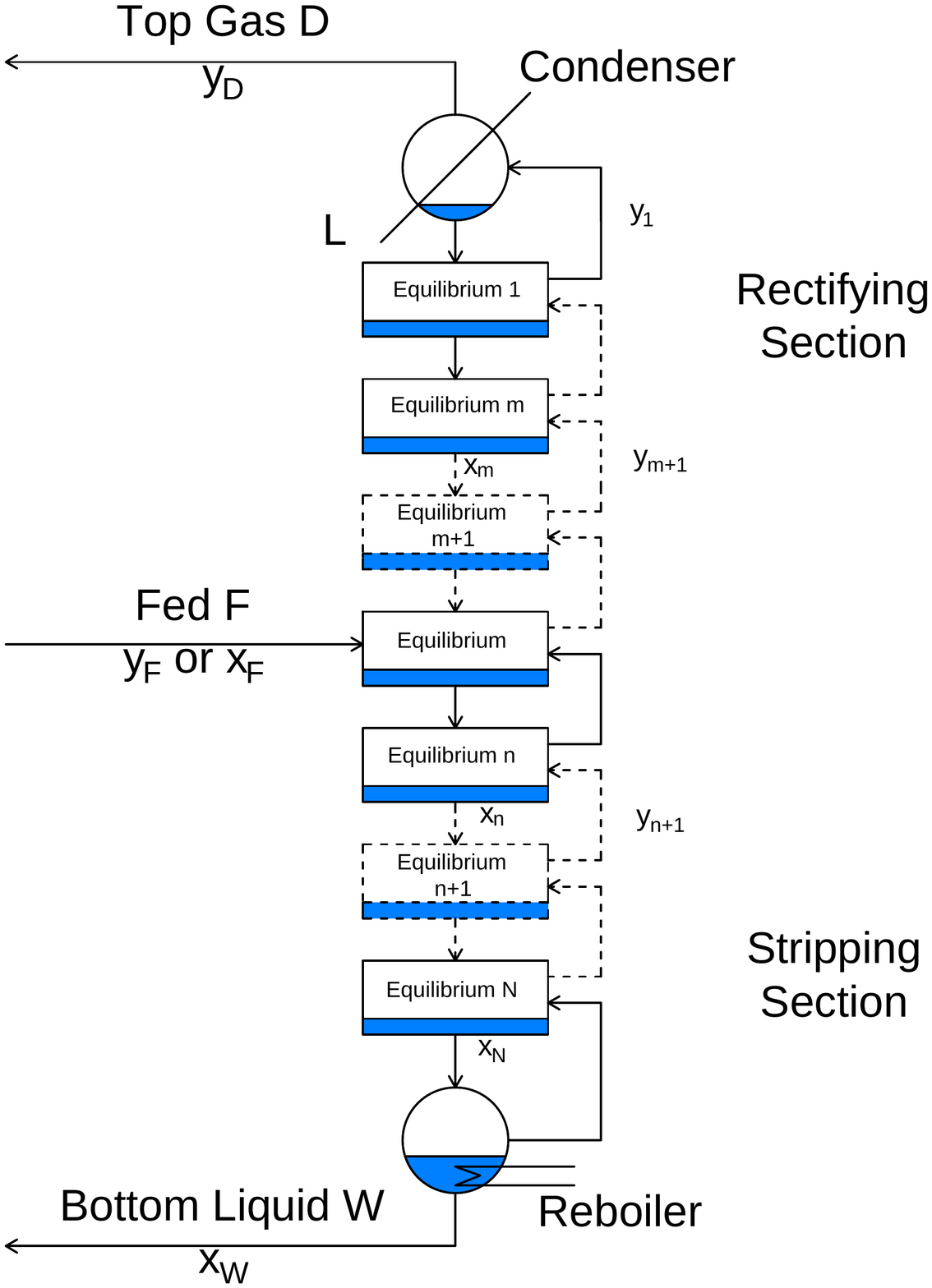}
\caption{\label{fig:2:1:2} Distillation method for the Kr-Xe and Rn-Xe separation, where $m$ and $n$ are the given theoretical stages in the rectifying section and stripping section, respectively.}
\end{figure}

The packed column can be segmented to several theoretical stages, in which the gas-liquid equilibrium is maintained. As shown in figure~\ref{fig:2:1:2}, the feeding flow is injected in the feeding stage with the flow rate $F$ and the concentration of volatile component $y_{F}$. Volatile component is enriched in the rectifying section above the feeding stage, which achieves a content $y_{D}$ eventually in the ceiling stage with an extracted flow rate of $D$. On the other side of the distillation column, the less volatile component is depleted stage by stage in the stripping section below the feeding stage, and the content declines to $x_{W}$ eventually in the base stage with an extracted flow rate of $W$.

Eq.~\eqref{eq:2:1:2} shows the conservation equations of the total mass flow and the concentration of volatile component, respectively.

\begin{subequations}\label{eq:2:1:2}
\begin{align}
\label{eq:2:1:2:1}
 F &= D + W \,,
\\
\label{eq:2:1:2:2}
 F \cdot y_{F} &= D \cdot y_{D} + W \cdot x_{W} \,.
\end{align}
\end{subequations}

Reflux ratio $R$ is another important factor, defined as the ratio of the reflux liquid mass flow $L$ and the extracted gas mass flow $D$ in the top condenser ($R=L/D$). It determines the liquid and vapor mass flow rate in the column, along with a liquid fraction $q$ in the feeding flow. During the Kr removal, the concentration of volatile component (Kr) in the rectifying section and stripping section can be expressed~\cite{textbook,XMass_kr1} by
\begin{subequations}\label{eq:2:1:3}
\begin{align}
\label{eq:2:1:3:1}
& \rm Rectifying\ Section : &y_{m+1}& = \frac{R}{R+1} \cdot x_{m} + \frac{1}{R+1} \cdot y_{D} \,,
\\ 
\label{eq:2:1:3:2}
& \rm Stripping\ Section :  &y_{n+1}& = \frac{RD+qF}{RD+qF-W} \cdot x_{n} - \frac{W}{RD+qF-W} \cdot x_{W} \,.
\end{align}
\end{subequations}
where $m$ and $n$ refer to the given theoretical stages in the rectifying section and the stripping section, respectively. In additional, $m+1$ or $n+1$ is the one right below $m$ or $n$. Similar relations hold for Rn removal as well.

\subsection{Column design method}
\label{sec:2:2}

The design of the column is driven primarily by the krypton removal requirement, as the Kr concentration has to be suppressed by seven orders of magnitude from commercial level (part-per-million or ppm) to 0.1~ppt. The design of the distillation system in this paper is based on the McCabe-Thiele (M-T) method~\cite{MT}.

\subsubsection{Reflux ratio and number of theoretical plates}
\label{sec:2:2:1}

The minimum reflux ratio, $R_{\rm min}$, corresponds to infinite number of stages of distillation. The $R_{\rm min}$ of the gas phase feeding ($q=0$) and the liquid phase feeding ($q=1$) are described in eq. \eqref{eq:2:2:1:1}~\cite{textbook}, in which the reduction factor has a condition of $1-y_{F}$, $1-y_{D} \sim 1$ and $y_{D} \gg y_{F}$ for the krypton removal.

\begin{subequations}\label{eq:2:2:1:1}
\begin{align}
\label{eq:2:2:1:1:1}
&  R_{\rm min}^{q=0} = \frac{\alpha}{ \alpha - 1 } \cdot \frac{y_{D}}{y_{F}} -1 \,,
\\
\label{eq:2:2:1:1:2}
& R_{\rm min}^{q=1} = \frac{1}{ \alpha - 1 } \cdot \frac{y_{D}}{y_{F}} \,.
\end{align}
\end{subequations}

The design of the distillation column is carried out based on the following assumptions and considerations.

\begin{itemize}
\item The M-T method is applicable under ultra-low concentration of the contaminant.
\item The krypton concentration in the original xenon is $y_{F}$ = 0.5 ppm ($5\times10^{-7}$~mol/mol) based on the National Standard of China.
\item For conservativeness, the designed target of the krypton concentration in the product xenon is $x_{W}$ = 0.01 ppt ($10^{-14}$~mol/mol).
\item The collection efficiency of xenon should be 99\%, which means $W^{\rm Kr} / F^{\rm Kr}$ = 0.99 and $D^{\rm Kr} / F^{\rm Kr}$ = 0.01, and $W^{\rm Rn} / F^{\rm Rn}$ = 0.01 and $D^{\rm Rn} / F^{\rm Rn}$ = 0.99.
\item The krypton removal speed should achieve 10~kg/h in order to process 6~tons of xenon in 30 days.
\item The maximum allowable height of the distillation column is 6 m considering the vertical space of CJPL.
\item Our air leak check sensitivity (combination of external and internal leaks) is better than $\sim$10$^{-9}$~Pa$\cdot$m$^{3}$/sec, which translates into an induced krypton concentration of $10^{-14}$~mol/day. Therefore, effects of the external leaks and internal krypton emanation is omitted from the design. 
\end{itemize}

Based on these considerations, the minimum reflux ratio determined from eq.~\eqref{eq:2:2:1:1} is $R_{\rm min}^{q=0} = 109$ for the gas phase feeding or $R_{\rm min}^{q=1} = 10.3$ for the liquid phase feeding. Since it is difficult to predict the liquid fraction in the feeding flow, both conditions have to be considered in the design.

For the packed column, the height of the theoretical plate (HETP) strongly depends on the gas loading, the type of mixture components and the packing structure. HETP = 35~cm is used for this design, the same value used in the first generation of the PandaX distillation system~\cite{panda2_dis1}. Note that this assumed value of HETP is conservative, given that similar packing materials and constructions from Sulzer~\cite{sulzer} have HETP values of a few cm under our maximum gas loading during the radon distillation (section~\ref{sec:5:2}). Under this height (35~cm), equilibrium is established for each theoretical stage, so the same HETP value is also applied for later calculation of the radon removal. For a 6~m total height, this corresponds to 17 ($N$ = 17) number of stages. 

\begin{figure}[htbp]
\centering
\includegraphics[width=.7\textwidth]{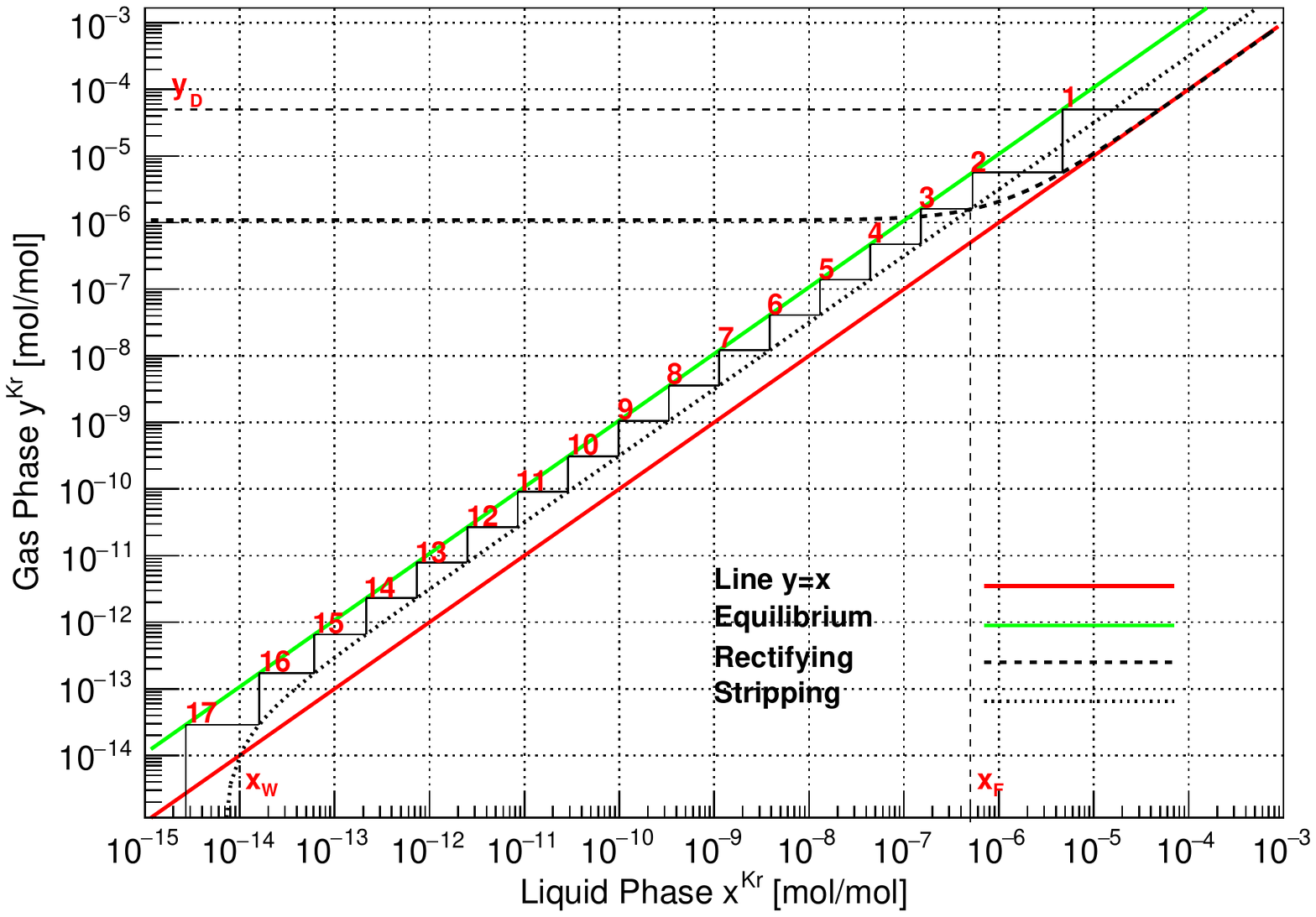}
\caption{\label{fig:2:2:1:1} M-T diagram for the krypton removal with liquid phase inlet, and the reflux ratio is 45.}
\end{figure}

In order to achieve a krypton contamination of less than 0.01~ppt, the reflux ratio for the liquid and gas feeding is required to be 45 and 145, respectively. The M-T diagram under the liquid phase feeding condition ($q=1$) with a reflux ratio 45 is shown in figure~\ref{fig:2:2:1:1}. The feeding point is at the theoretical stage 3 (stage 1 is the top stage in height conventionally). For the gas phase feeding condition, the diagram changes slightly and the feeding point is the same.
Because that the gas-liquid equilibrium curve is linear in ultra-low concentration region ($x_{F} \ll 1$), the same reduction factor can be achieved independent of the feeding krypton concentration.

\begin{figure}[htbp]
\centering 
\includegraphics[width=.7\textwidth]{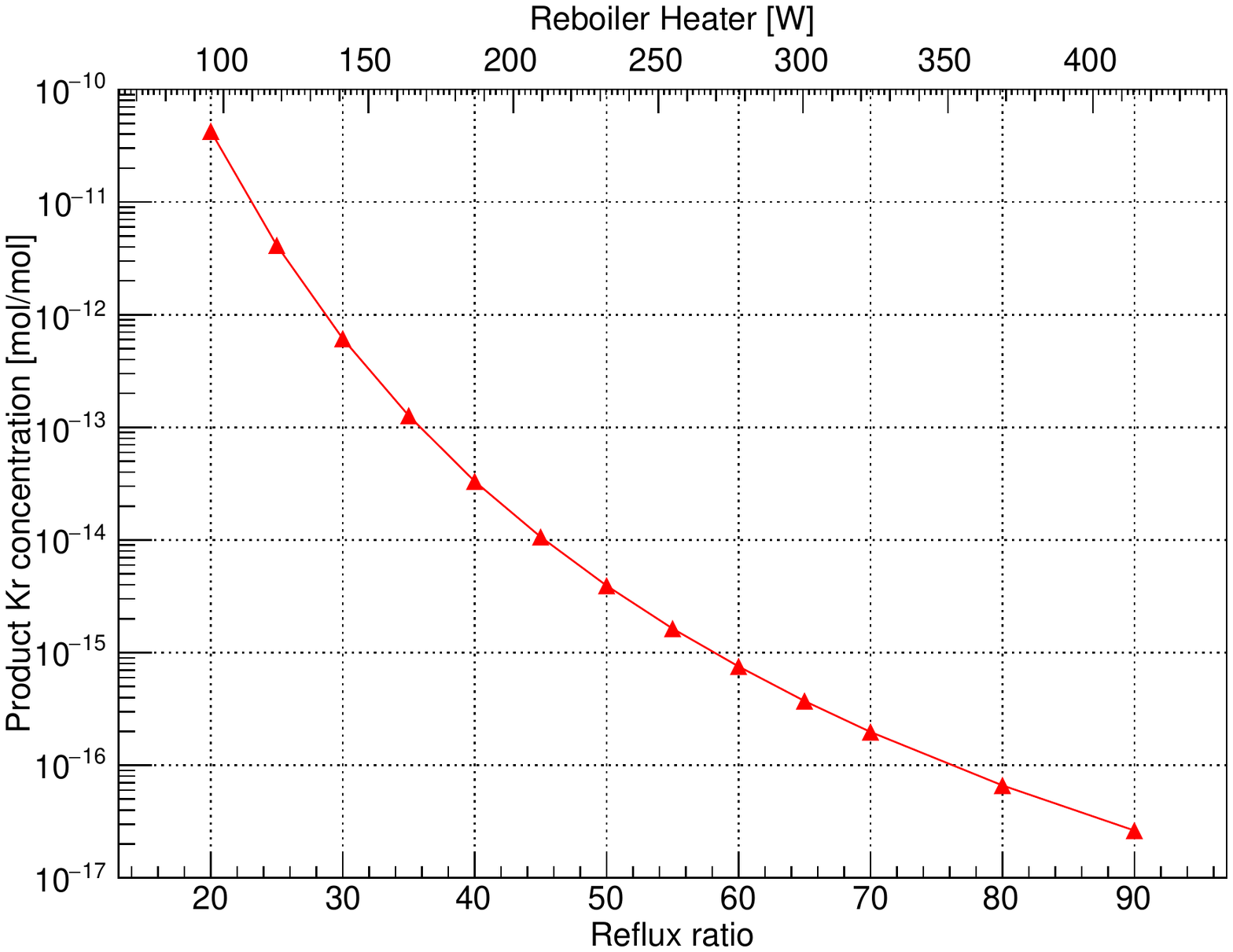}
\caption{\label{fig:2:2:1:2} Product Kr concentration with different reflux ratio. The upper x-axis corresponds to the reboiling heating power for the 17.7 kg/h feeding flow.}
\end{figure}

We repeat the M-T method for different reflux ratio, and estimate the product krypton concentration. The results are summarized in figure~\ref{fig:2:2:1:2}. Larger reflux ratio leads to less krypton concentration, but it also requires more reboiler heating power which makes the system more prone to flooding. 

\subsubsection{Superficial velocity and the column diameter}
\label{sec:2:2:2}

\begin{figure}[htbp]
\centering
\includegraphics[width=.35\textwidth]{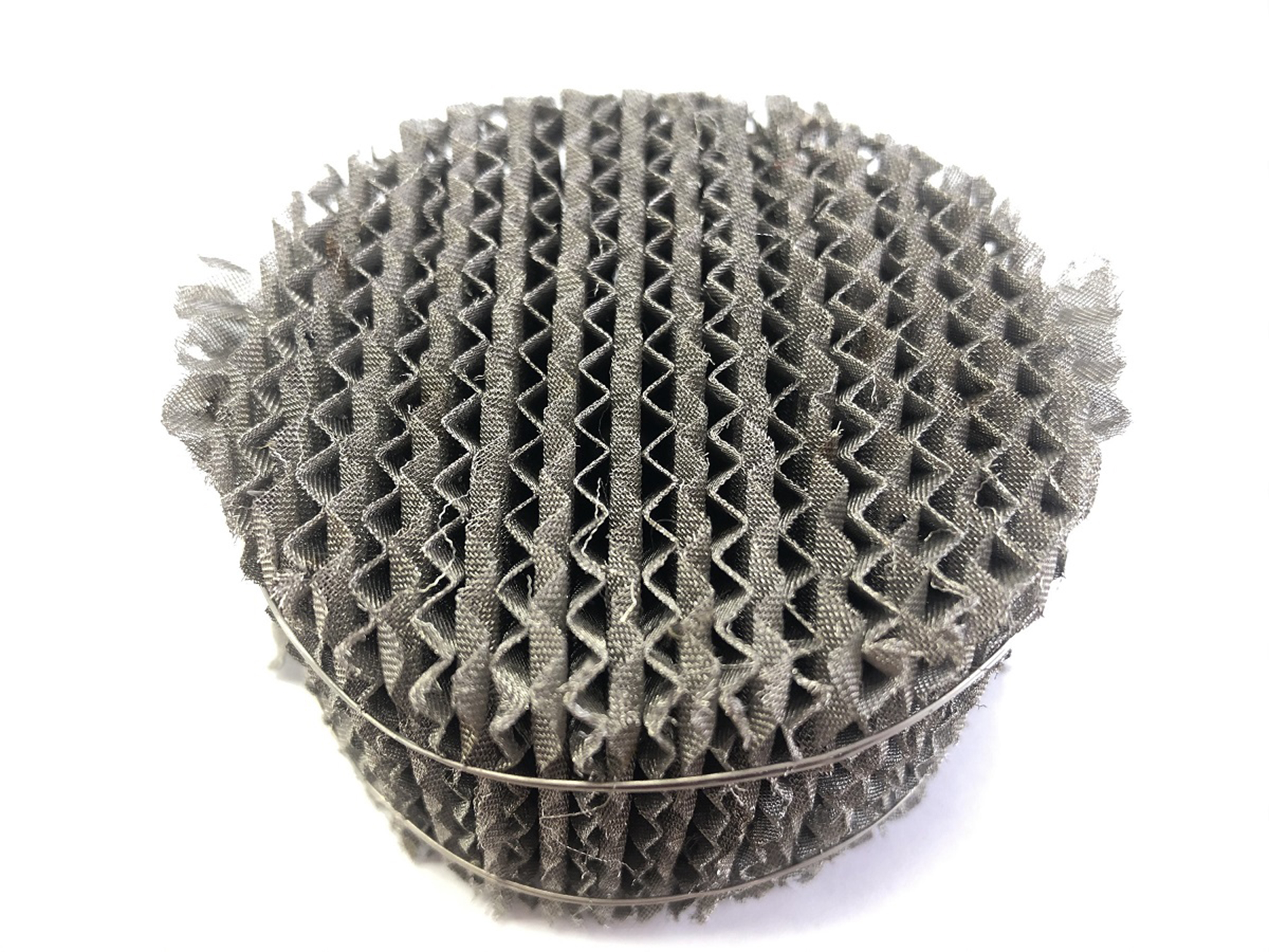}
\caption{\label{fig:The structured packing PACK-13C} The structured packing PACK-13C.}
\end{figure}

Flooding is a typical incident during the distillation operation which causes gaseous pressure rising continuously with liquid stuck on the packing, mostly due to non-condensable gas or particulates in the feeding flow. Larger column diameter can lower the possibility of flooding and 
the Bain-Hougen correlation~\cite{BH_co} is usually applied for the calculation of the flooding velocity and the column diameter for the structured packing column. The equation is shown in eq.~\eqref{eq:2:2:2:1}. For simplicity, we only consider the rectifying section of the tower and $q=0$, since these give the largest gas loading compared to other conditions.

\begin{equation}
\label{eq:2:2:2:1}
\begin{split}
{\rm log} \left( \frac{u_{f}^{2}}{g} \frac{\sigma}{\varepsilon ^{3}} \frac{\rho _{V}}{\rho _{L}} \mu _{L}^{0.2} \right) = P - N \left( \frac{Q_{L}}{Q_{V}} \right) ^{1/4} \left( \frac{\rho _{V}}{\rho _{L}} \right) ^{1/8} \,.
\end{split}
\end{equation}
In this equation, $u_{f}$ is the flooding gaseous velocity, $g$ is the acceleration of gravity, $\mu _{L}$ is the viscosity of the liquid phase, $Q_{L}=0.01F \cdot R$ and $Q_{V} = 0.01F \cdot (R+1)$ is the mass flow rate for the liquid and gas phase, respectively. $\rho _{V}$ and $\rho_{L}$ are the gas and liquid xenon density, respectively. The same packing PACK-13C~\cite{packing} used in the first generation of the PandaX column is also adopted here (figure~\ref{fig:The structured packing PACK-13C}). In eq.~\eqref{eq:2:2:2:1}, the corresponding packing constants are $P$ = 0.73 and $N$ = 2.28, and $\sigma$ = 1135 m$^{2}$/m$^{3}$ and $ \varepsilon$ = 0.77 m$^{3}$/m$^{3}$ are the specific surface area and the voidage of the packing, respectively.

The superficial gaseous velocity $u = 0.5 \cdot u_{f}$ is used in the column design, in which the flooding gaseous velocity $u_{f}$ is obtained from eq.~\eqref{eq:2:2:2:1}. $u$ is 0.25~m/s. Based on eq.~\eqref{eq:2:2:2:2}~\cite{textbook}, the minimum value of $d$ can be determined to be 34~mm.
\begin{equation}
\label{eq:2:2:2:2}
\begin{split}
d = \sqrt{ \frac{4 \cdot Q_{V}}{3600 \pi \cdot u \cdot \rho_{V}} } \,.
\end{split}
\end{equation}

On the other hand, larger diameter leads to losses in liquid-gas exchange efficiency, so liquid sprayed density $U$, defined in eq.~\eqref{eq:2:2:2:3}~\cite{textbook}, an index to describe the influence of the liquid infiltration of the packing, has to be considered. 
\begin{equation}
\label{eq:2:2:2:3}
\begin{split}
U = \frac{Q_{L}}{0.785 \cdot d^{2} \cdot \rho _{L}} \,.
\end{split}
\end{equation}
Empirically, $U$ should be larger than $U_{\rm min}$ = 0.24~m$^{3}$/(m$^{2}\cdot$h) for the structured packing column. Therefore, the maximum column diameter is 164~mm. In reality, $d=125$~mm is chosen as the diameter of the distillation column, which can easily fit into commercial stainless steel tubes.

\subsection{Radon removal}
\label{sec:2:3}

In contrast to krypton removal, xenon is the more volatile component compared to radon. Therefore under radon removal operation,
the product xenon is enriched in the condenser at the top of the distillation tower. 
The radon operation parameters can be obtained for this tower. We only consider the gaseous phase feeding condition ($q=0$)
due to the consideration of the heat exchanger condition in figure~\ref{fig:3:1} (see section~\ref{sec:3}).

The collection efficiency is set at 99\% for the radon removal. The equation of the minimum reflux ratio is shown in eq.~\eqref{eq:2:3:1}~\cite{textbook},
\begin{equation}
\label{eq:2:3:1}
\begin{split}
R_{\rm min}^{\rm Rn} = \frac{1}{\alpha^{\rm Rn} -1} \left( \frac{\alpha^{\rm Rn} \cdot y_{D}}{y_{F}} - \frac{1-y_{D}}{1-y_{F}} \right) -1 \sim \frac{1}{\alpha^{\rm Rn} -1} \,,
\end{split}
\end{equation}
 in which we take $y_{D},y_{F} \sim 1$ and $1-y_{D} \ll 1-y_{F}$, since in this case the radon concentration is much lower than xenon, and the distillation reduction factor should be much larger than 1. Given $\alpha^{\rm Rn} = 11.3$ (section~\ref{sec:2:1}), the minimum reflux ratio is $R_{\rm min}$ = 0.097.
 %0.078
 Therefore 
 a reflux ratio $R^{\rm Rn}$ of 0.15 is chosen for the Rn distillation. Based on eq.~\eqref{eq:2:2:2:3} and $Q_{L}^{\rm Rn}=0.99 F \cdot R^{\rm Rn}$, the resulting minimum flow rate is 56.5~kg/h.

The radon concentration in PandaX-II, $y^{\rm Rn}$ = 3.3$\times$10$^{-24}$~mol/mol, is used as the feeding condition for the Rn removal calculation. Additional Rn emanation comes from the tower itself, as the packing contains uranium impurities. 
In order to control the uranium contamination on the surface, radioactivity level of the packings after various cleaning procedures are measured by a high purity germanium detector (HPGe)~\cite{HPGe}. A mixture of 13\% HNO$_{3}$ and 0.2\% HF is selected as the etching solution. The total mass of the packing is about 25 kg. After etching, the $^{222}$Rn activity in the packing is measured to 
be <6~mBq/kg, below the sensitivity of the HPGe detector.
Under the secular equilibrium assumption, even if 50\% $^{222}$Rn atoms emanate from the packing, the total emanation rate is 1.4$\times10^{-24}$ mol/s. The distillation performance is evaluated with and without the assumption of this tower emanation in the next two sections.

\subsubsection{Radon reduction without tower emanation}
\label{sec:2:3:1}
\begin{figure}[htbp]
\centering
\includegraphics[width=.7\textwidth]{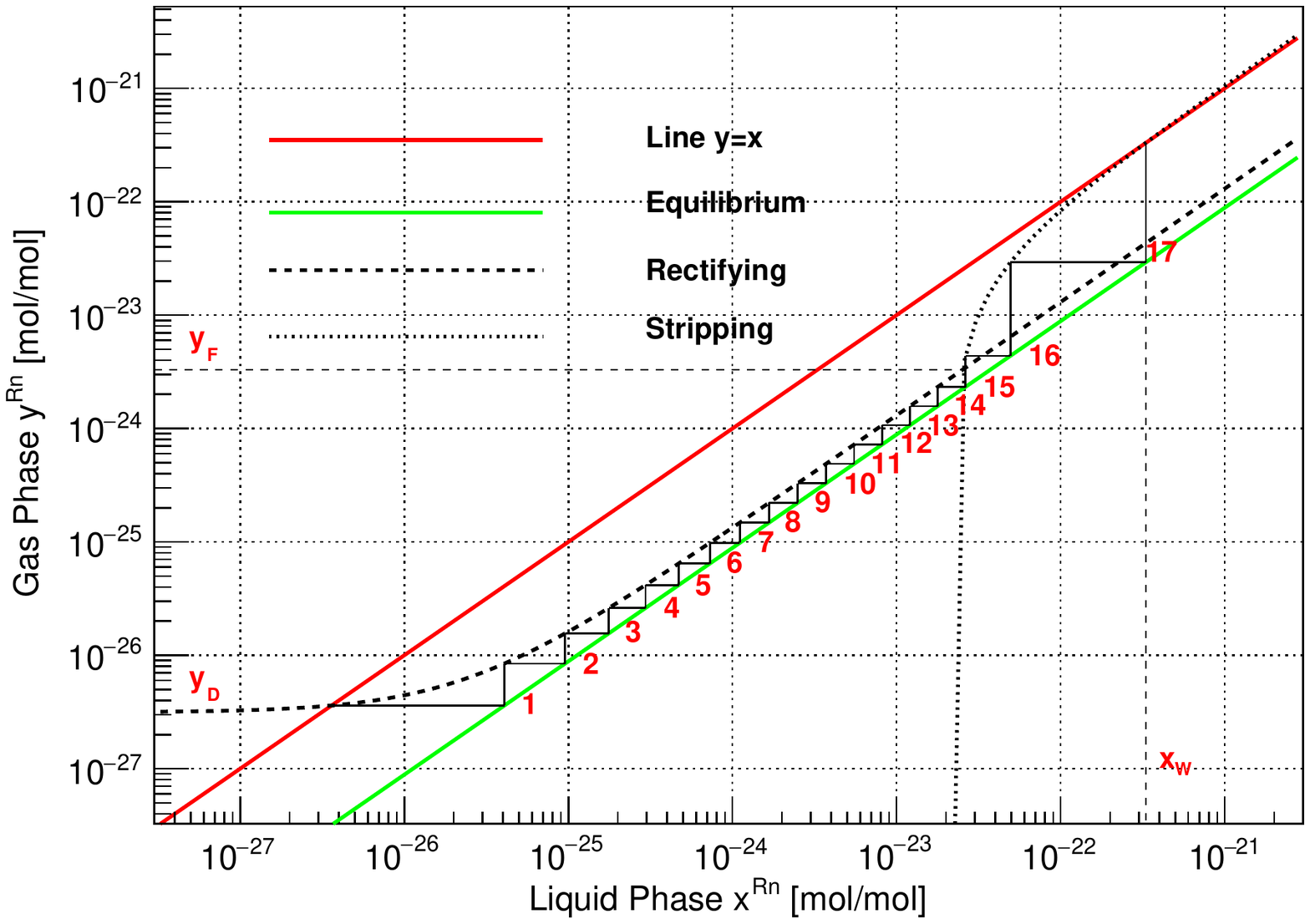}
\caption{\label{fig:2:3:1:1} M-T diagram for the radon removal without the tower emanation, and the reflux ratio is 0.15.}
\end{figure}

Given the equilibrium equation in eq.~\eqref{eq:2:1:2}~\cite{textbook}, the radon concentration in the xenon-radon mixture can be written as
\begin{equation}
\label{eq:2:3:1:1}
\begin{split}
 y^{\rm Rn} = 1 - y^{\rm Xe} = 1 - \frac{\alpha^{\rm Rn} \cdot \left(1-x^{\rm Rn} \right)}{1 + \left( \alpha^{\rm Rn} - 1 \right) \cdot \left(1-x^{\rm Rn} \right) } \sim \frac{x^{\rm Rn}}{\alpha^{\rm Rn}} \,.
\end{split}
\end{equation}
Similarly, eq. \eqref{eq:2:1:3} can also be applied for the radon removal process, where all parameters are for Rn. Combining all these, the M-T diagram is shown in figure~\ref{fig:2:3:1:1}. The resulting reduction factor for the radon removal is $Re_{\rm dis}^{\rm Rn} \sim 10^{3}$. The theoretical stage number of the feeding point is 15 (close to the bottom), which is opposite from the krypton removal.

\subsubsection{Radon reduction with tower emanation}
\label{sec:2:3:2}

In this section, all radon emanated from each theoretical plate is assumed as a constant emanation rate $E$ with 
$1.4\times10^{-24}/17 = 8.2\times10^{-26}$~mol/s/stage.
The equilibrium equation is the same as the eq. \eqref{eq:2:3:1:1}. After the inclusion of the emanation, the process equations derived from the conservation of matter, eqs.~\eqref{eq:2:1:2} and ~\eqref{eq:2:1:3}, can now be written as
\begin{subequations}\label{eq:2:3:2:1}
\begin{align}
\label{eq:2:3:2:1:1}
& \rm Conservation\ of\ radon : &F \cdot y_{F}^{\rm Rn}& = D \cdot y_{D}^{\rm Rn} + W \cdot x_{W}^{\rm Rn} -N \cdot E \,,
\\
\label{eq:2:3:2:1:2}
& \rm Rectifying\ Section : &y_{m+1}^{\rm Rn}& = \frac{ RD \cdot x_{m}^{\rm Rn} + D \cdot y_{D}^{\rm Rn} - m \cdot E}{RD+D} \,, 
\\
\label{eq:2:3:2:1:3}
& \rm Stripping\ Section :  &y_{n+1}^{\rm Rn}& = \frac{RD \cdot x_{n}^{\rm Rn} - W \cdot x_{W} + \left( N-n-1 \right) \cdot E} {RD-W} \,,
\end{align}
\end{subequations}

where $N=17$ is the total number of stages.

Due to the emanation effect, traditional M-T method is not directly applicable. An iterative method is used to obtain the product radon 
concentration $y_D$. We first fix the feeding stage number, and assume a value for $y_D$. Then $x_{W}$ can be obtained 
from eq.~\eqref{eq:2:3:2:1:1}.  Then using eq.~\eqref{eq:2:3:2:1:2} and 
eq.~\eqref{eq:2:3:2:1:3}, the radon concentration at the feeding stage can be calculated, respectively, from 
the top and bottom of the tower. $y_D$ is adjusted until the two results match. This procedure can be repeated for different feeding stages, resulting into an optimal feeding stage at number 4.

\begin{figure}[htbp]
\centering
\includegraphics[width=.7\textwidth]{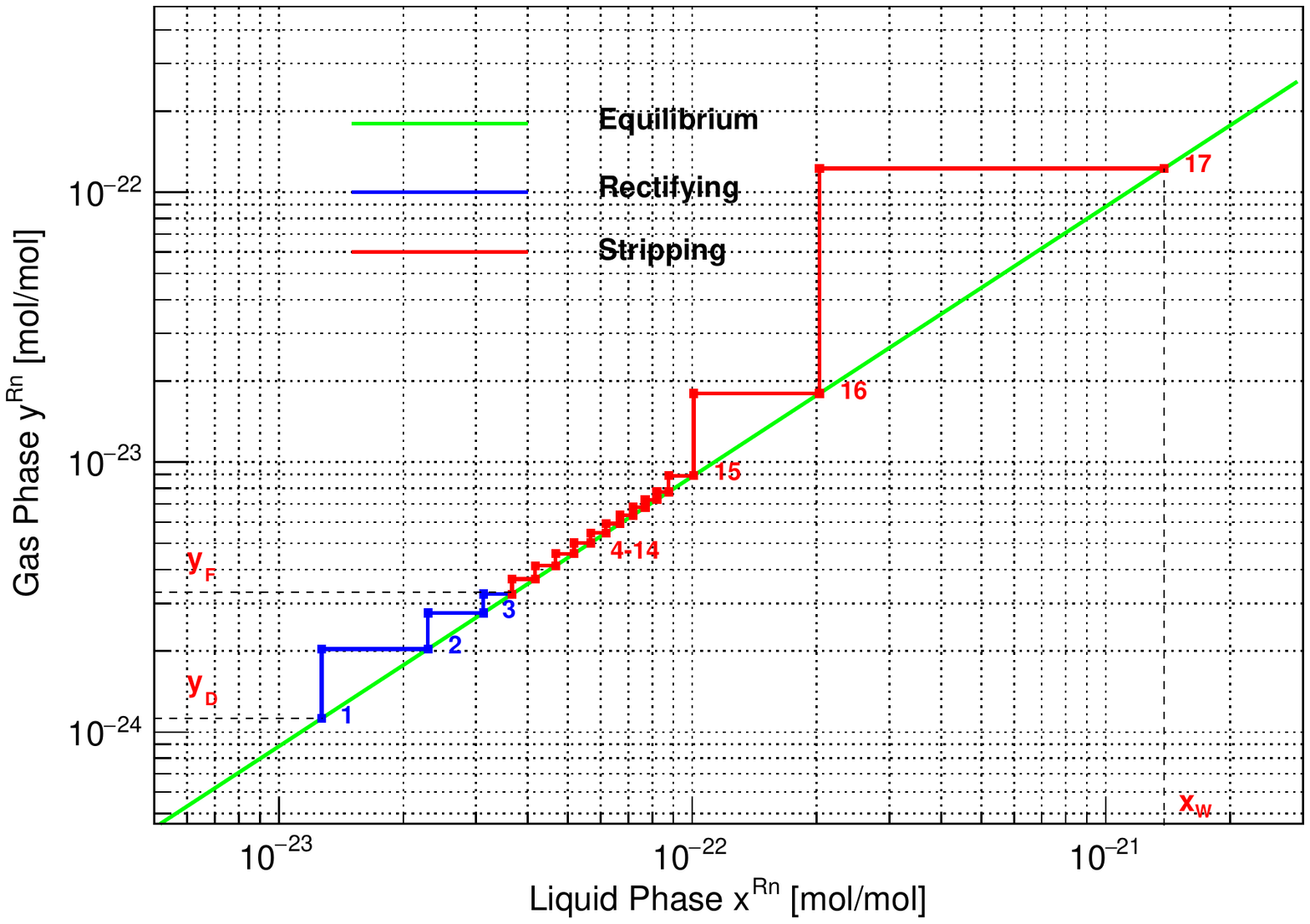}
\caption{\label{fig:2:3:2} Radon concentration diagram with tower emanation, and the reflux ratio is 0.15.}
\end{figure}

The corresponding radon concentration diagram is shown in figure~\ref{fig:2:3:2}. For the rectifying section, by combining equations \eqref{eq:2:3:2:1:2} and \eqref{eq:2:3:1:1}, we have

\begin{equation}
\label{eq:2:3:2:3}
y_{m+1} - y_{m} = \left( \frac{R}{R+1} - \frac{1}{\alpha} \right) x_{m} - m \cdot \frac{E}{D \cdot (R+1)} + \frac{y_{D}}{R+1}\,,
\end{equation}
The $E$ term clearly shows that each plate is affected by the emanated radon from all plates above (enrichment effect), which leads to a weakening reduction of the radon for larger plate number. The radon reduction factor is $Re_{\rm dis}^{\rm Rn} \sim 3$, and less radon emanation rate will increase the reduction factor.

\subsubsection{Estimate radon reduction factor in the PandaX-4T detector}
\label{sec:2:3:3}

Similar to the tower emanation, a realistic detector has a constant radon emanation rate. While the reduction factor for a standalone 
tower is large, as described in the previous section, the final $^{222}$Rn concentration in the detector is a dynamic balance of the continuous emanation, decay ($\lambda_{\rm Rn^{222}} = 2.1 \times 10^{-6}$ s$^{-1}$), and the distillation reduction. The distillation system can not reduce the concentration of $^{220}$Rn in the detector, for its half-life is much shorter (55 s). 

In order to transfer xenon from the detector to the distillation system, the online distillation will be connected to the circulation system of the detector. Along the loop, three types of radon emanation sources are identified, similar to those in ref.~\cite{xenon_rn}. Type 1 source ($k_{1}$) is located downstream of the distillation system, including the detector materials and the returned circulation loop upstream of the detector. Type 2 source ($k_{2}$) is located upstream of the distillation system, which can be effectively removed by the distillation. Type 3 source ($k_{3}$) is the emanation of the tower discussed in previous section. For infinitely long time, the number of radon atoms in the detector with and without the distillation Rn reduction factor $Re^{\rm Rn}_{\rm dis}$ are given as
\begin{subequations}\label{eq:2:3:3:1}
\begin{align}
\label{eq:2:3:3:1:1}
N_{\rm without} &\overset{t \to \infty }{=} \frac{k_{1} + k_{2}}{\lambda _{\rm Rn}} \,,
\\
\label{eq:2:3:3:1:2}
N_{\rm with} &\overset{t \to \infty }{=} \frac{k_{1} + k_{2}/Re_{\rm dis}^{\rm Rn}}{ \lambda _{\rm Rn} + f \cdot \left( 1 - 1/Re_{\rm dis}^{\rm Rn} \right) } \,. 
\end{align}
\end{subequations}
In the above relation, $f$ is the ratio of the distillation mass flow rate and the total xenon mass in the detector, with a value $2.6 \times 10^{-6}$ s$^{-1}$ given the flow rate of 56.5~kg/h and the total 6 tons of xenon mass. $k_1$ is assumed to be the same as the PandaX-II measurement (32~$\mu$Bq/kg$\times$6~ton = 192~mBq). Based on an offline measurement with the entire circulation loop, $k_2$ is measured to be 6~mBq. In addition, the type 2 rate will be further suppressed by the distillation before entering the detector. So the radon contribution from the type 2 source can be neglected.
As a result, when reaching the dynamic balance, the radon reduction factor is expressed as
\begin{equation}
\label{eq:2:3:3:2}
\begin{split}
Re_{\rm Detector}^{\rm Rn} = \frac{\lambda _{\rm Rn}}{ \lambda _{\rm Rn} + f \cdot \left( 1 - 1/Re_{\rm dis}^{\rm Rn} \right)} \,.
\end{split}
\end{equation}
For the radon reduction factor $Re_{\rm detector}^{\rm Rn}$ varying with the distillation radon reduction factor $Re_{\rm dis}^{\rm Rn}$, which is related to the tower radon emanation rate and the feeding radon rate, $Re_{\rm detector}^{\rm Rn}$ is 1.83 or 2.24 with or without the tower radon emanation effect. Clearly, in this case the operation mass flow rate is more relevant to the radon reduction factor in the detector than the distillation reduction factor.

\section{Construction of the tower}
\label{sec:3}

\begin{figure}[htbp]
\centering 
\includegraphics[width=.99\textwidth]{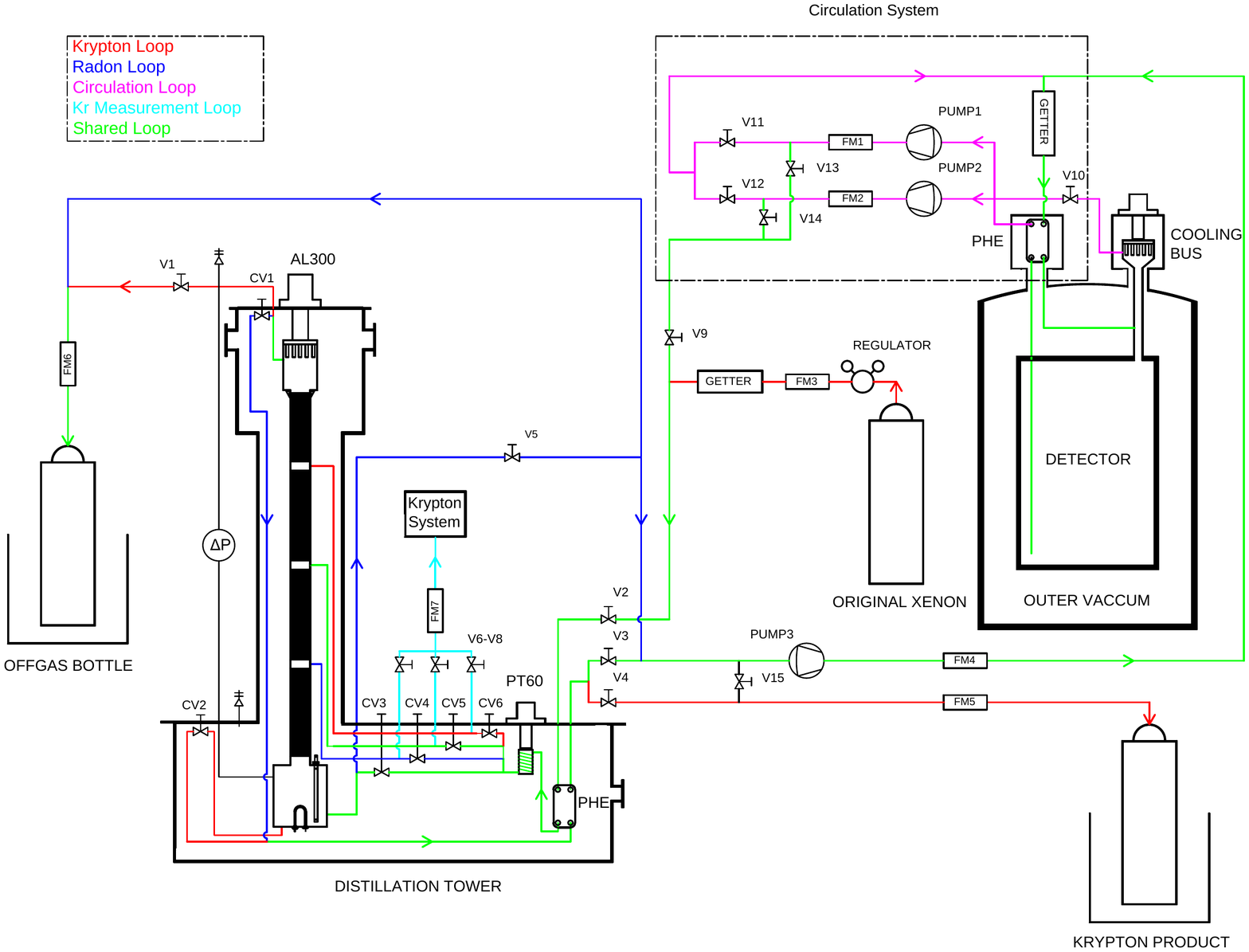}
\caption{\label{fig:3:1} The P\&I diagram for the PandaX-4T distillation system. The red and blue lines are for the krypton and radon removal operations, respectively, and the green lines are shared for both. The purple and cyan lines are the circulation loop and the krypton measurement system loop, respectively.}
\end{figure}

In the previous section, the basic dimensions and requirements of the distillation tower have been settled. A simplified piping and instrument (P\&I) diagram of the tower and its interfaces with the PandaX-4T detector 
is illustrated in figure~\ref{fig:3:1}. 
The tower has three feeding positions with the consideration of three operation modes and the actual installation condition, 1.5~m, 2.8~m and 4.4~m, measuring from the top. 
In this section, we shall discuss its key components of the tower under pragmatic considerations.

Since the feeding flow is gaseous xenon of 178 K, the power requirements of the reboiler and the condenser can be estimated as

\begin{subequations}\label{eq:3:1}
\begin{align}
\label{eq:3:1:1}
Q_{\rm Reb} & = \left( L+F\cdot q - W \right) \cdot h = \left[ \left( R+1 \right) \cdot D - (1-q) \cdot F \right] \cdot h \,,
\\
\label{eq:3:1:2}
Q_{\rm Con} & = L \cdot h= R \cdot D \cdot h \,,
\end{align}
\end{subequations}

in which $h=92.6$~kJ/kg is the xenon latent heat of liquefaction. Therefore, the power requirement of the condenser and the reboiler are about 370~W and 200~W, respectively, taking the maximum need for radon and krypton. 
A GM cryocooler (Cryomech, AL300)~\cite{cryomech} which can supply up to $\sim$400~W cooling power at 178~K is chosen for the condenser. A plate heat exchanger (PHE) (KAORI, K70)~\cite{KAORI} and a pulse tube refrigerator (Cryomech, PT60)~\cite{cryomech} are used for the pre-cooling of the feeding xenon gas. 

The pressure drop $\Delta P$ between the condenser and the reboiler is an important indicator of the tower performance. A differential pressure gauge (MKS, 226A)~\cite{MKS} is used to monitor $\Delta P$ during the operation. Three capacitance liquid level meters are installed in the reboiler. The volume of the reboiler is 30 liters, which can store 90~kg liquid xenon. Three heating rods each with 180~W maximum heating power are used to evaporate xenon from the reboiler.

There are several sampling ports on the tower, connected with a high sensitivity krypton measurement system directly, to assay the gas purity in the feeding points, the condenser and the reboiler.

\begin{figure}[htbp]
\centering
\includegraphics[width=.6\textwidth]{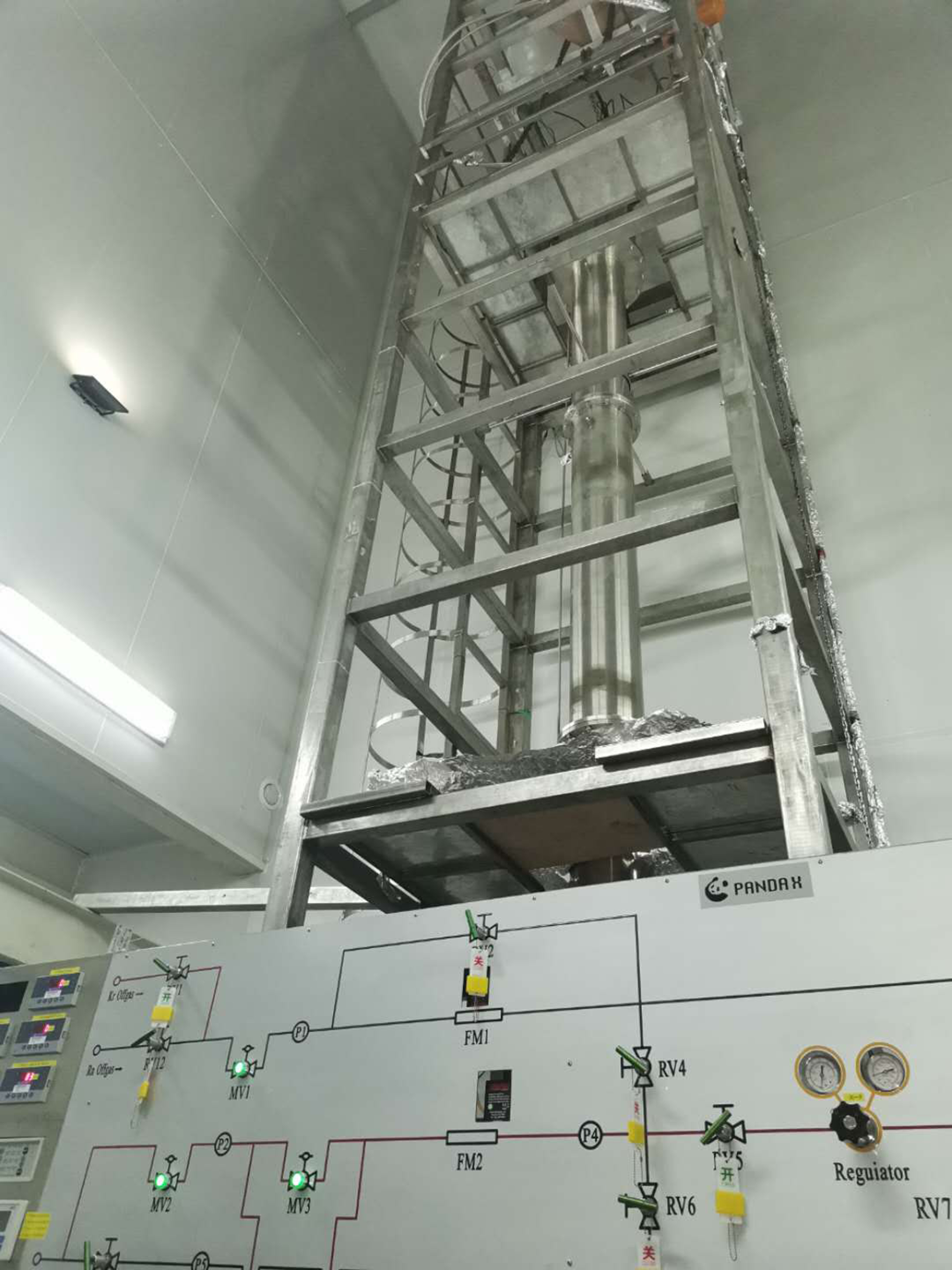}
\caption{\label{fig:3:2} The photo of PandaX-4T cryogenic distillation system for krypton and radon removal.}
\end{figure}

A photo of the as-built distillation tower system is shown in figure~\ref{fig:3:2}. 

\section{Operation modes of the distillation tower}
\label{sec:4}

By design, the distillation tower can be operated under the offline krypton, online krypton, and the online radon removal modes, 
where the word offline or online refers to whether the tower is operated together with the circulation of the xenon detector. 
The offline krypton removal should be performed first before the xenon filling into the detector. After filling, to remove potential krypton due to initial outgassing or external leak, online krypton purification is envisioned to suppress krypton to the desired level. 
Since radon emanation is continuous, the online radon removal should be uninterrupted during the detector data taking. As the product 
xenon is extracted from different positions from the tower, the krypton and radon removal can not be realized at the same time.

During the offline krypton removal operation, the original xenon comes from the high pressure bottles, and the flow rate is controlled by the regulator and the mass flow controller. On the other hand, for the online krypton removal, the xenon comes from the detector circulation loop, and the mass flow rate is kept by the circulation diaphragm pump. The feeding xenon flow is fed into the column at the top feeding position after the pre-cooling. The product xenon is extracted from the reboiler at the bottom of the distillation tower, and will go through the PHE into the detector during the online mode, or 40 liter aluminum bottles cooled by liquid nitrogen during the offline mode.
Offgas is extracted from the condenser at the top of the tower and stored in an aluminum bottle cooled by liquid nitrogen as well.

During the online radon removal process, the gaseous product xenon is produced in the condenser, and goes into the PHE. Similar to the online krypton removal process, the flow of the feeding xenon and product xenon is achieved by the circulation system. The offgas, enriched with radon, is stored in the reboiler, and can be extracted. Since radon decays, the extracted xenon can be put back into the detector after some period.

\section{Commissioning operation and optimization}
\label{sec:5}

Ten days of the distillation tower commissioning test was completed without flooding. In order to simulate the actual distillation-detector working condition, the self circulation of the distillation system was achieved with one diaphragm pump and a buffer. During the commissioning test, the krypton offline and radon online procedures were successfully operated with the designed mass flow rate and reboiler heating power. Different mass flow rate and reboiler heating power were studied to optimize the running conditions for the tower.

\begin{figure}[htbp]
\centering 
\subfigure[The total reflux period]{
\label{fig:5:1:a}
\begin{minipage}[t]{0.45\textwidth}
\centering
\includegraphics[width=6.5cm]{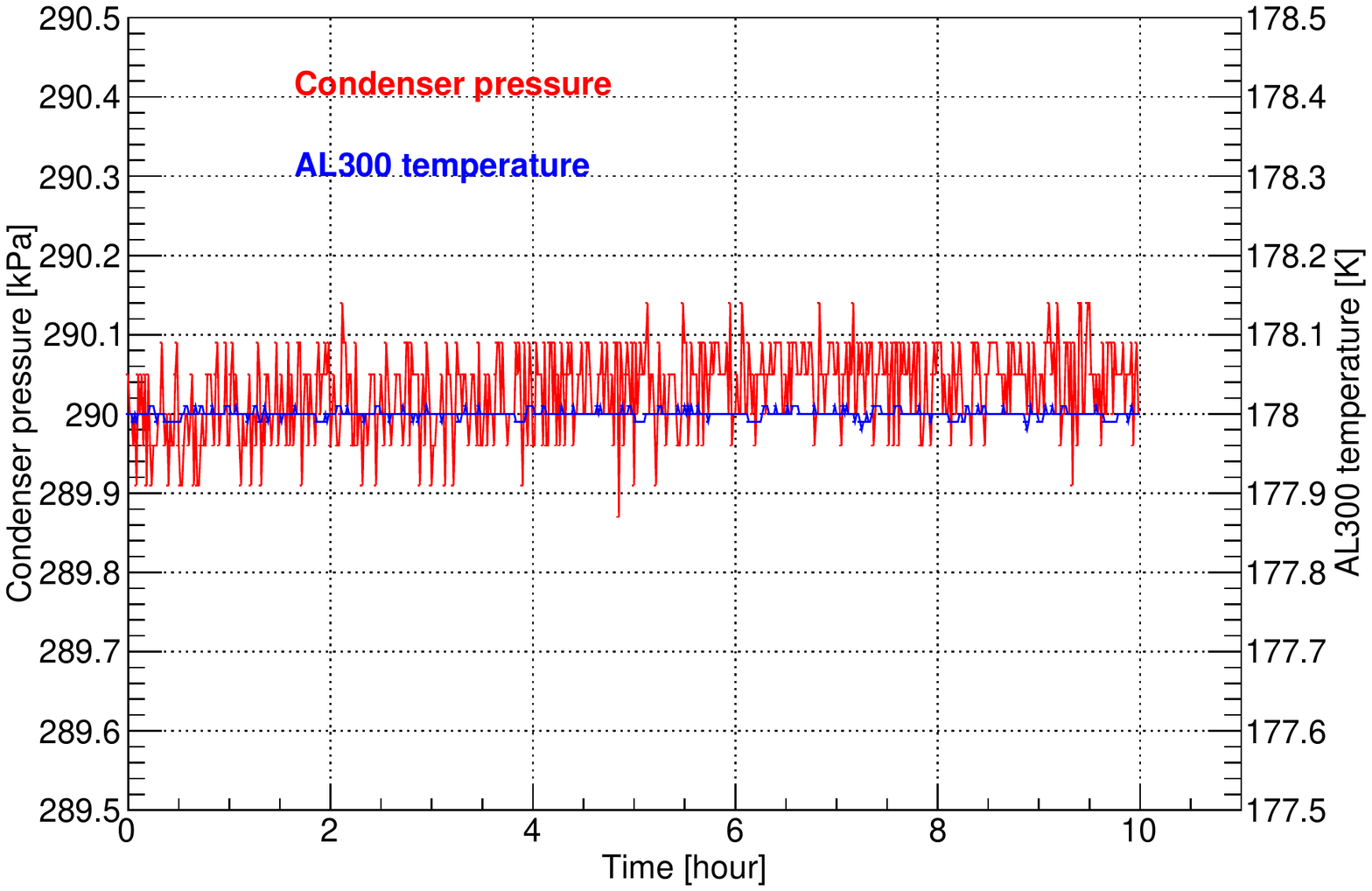}
\end{minipage}
}
\subfigure[The radon removal period]{
\label{fig:5:1:b}
\begin{minipage}[t]{0.45\textwidth}
\centering
\includegraphics[width=6.5cm]{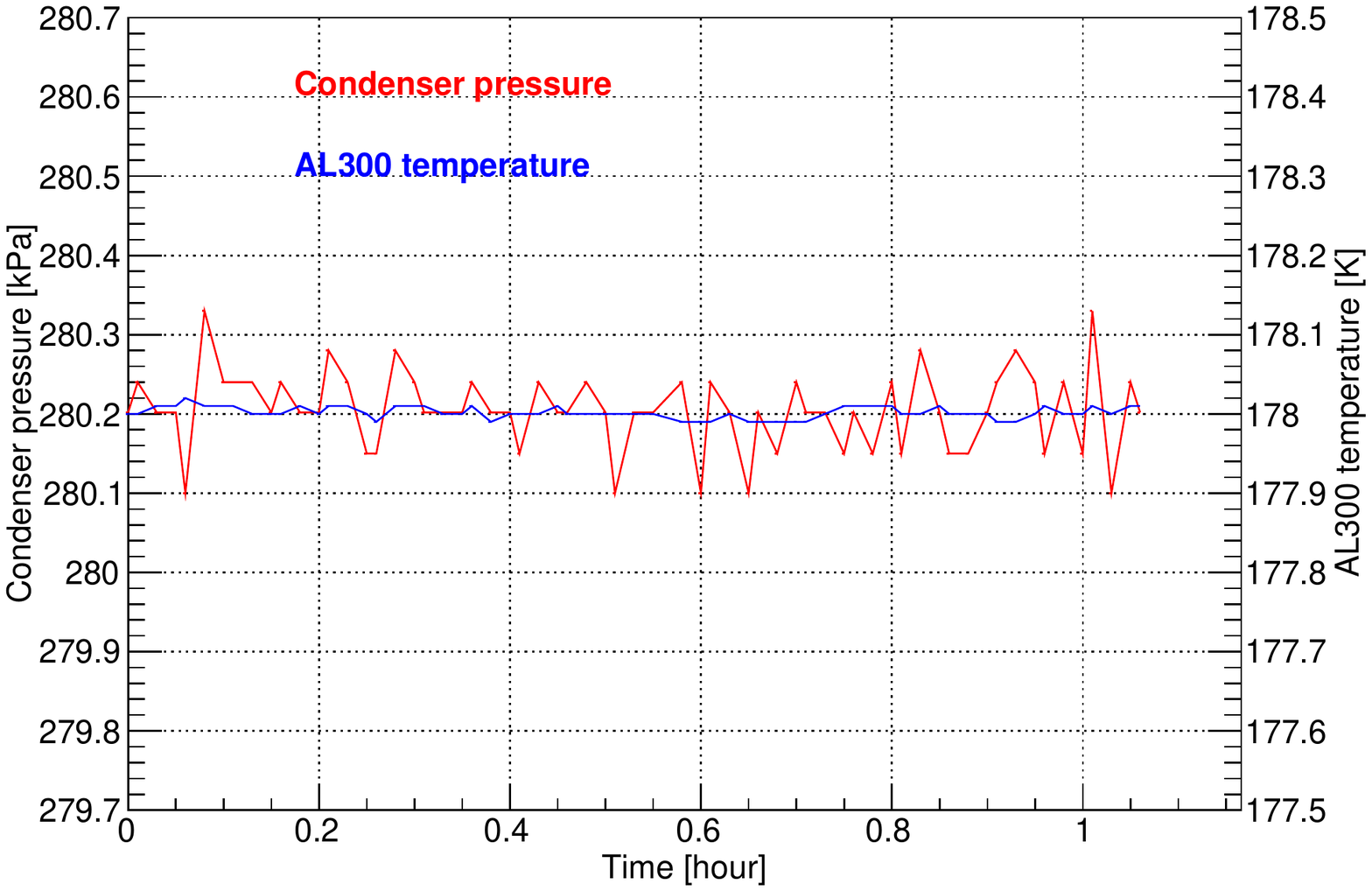}
\end{minipage}
}
\caption{ \label{fig:5:1} Temperature and pressure variations during the commissioning test.}
\end{figure}

Figure~\ref{fig:5:1} shows the AL300 temperature and the condenser pressure fluctuations during the commissioning test. The tower operation was quite stable, with a temperature and pressure of the condenser varying by less than 0.04~K and 0.4~kPa during the total reflux period and the radon operation period, respectively shown in figure~\ref{fig:5:1:a} and figure~\ref{fig:5:1:b}. And the liquid level in the reboiler varied by less than 10~mm. 

\subsection{Efficiency of the heat exchanger}
\label{sec:5:1}

\begin{figure}[htbp]
\centering 
\includegraphics[width=.7\textwidth]{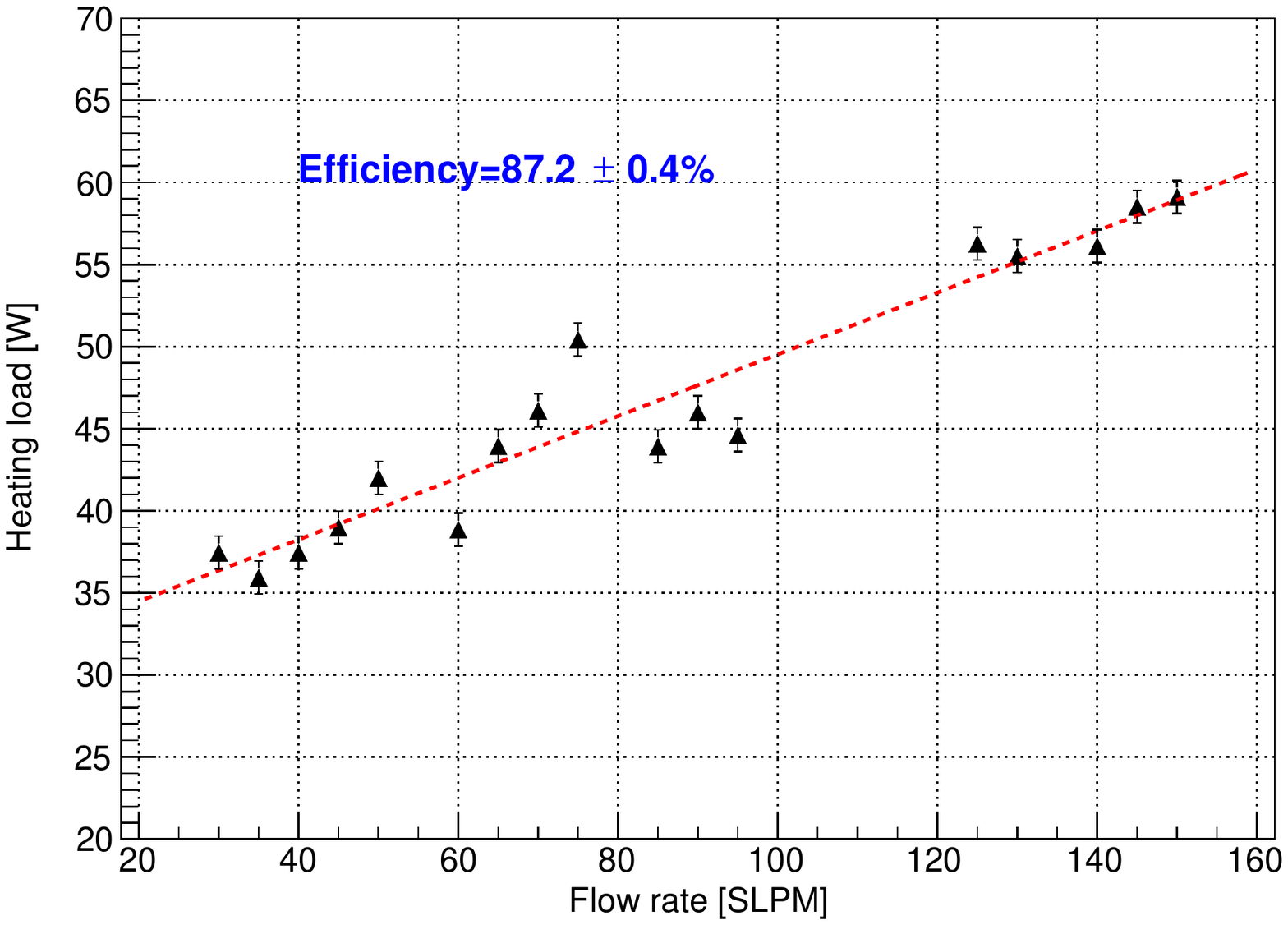}
\caption{\label{fig:5:1:1} Heating load with inlet flow during Rn removal process.}
\end{figure}

Different liquid fraction of the feeding xenon will affect the product concentration, so the efficiency of the pre-cooling PHE is important. For the radon removal process, since the product xenon is extracted from the top of the tower in the gaseous phase, as gas-gas heat exchange happens inside the PHE, the feeding xenon will remain in gaseous form after the PHE before entering the tower. Figure~\ref{fig:5:1:1} shows the AL300 heating load versus different feeding flow rate during the radon removal process. The gas-gas heat exchange efficiency is determined to be 87.2\%, using the temperature drop from 293~K to 178~K for the gaseous xenon. 

For the krypton removal, on the other hand, since the product xenon is extracted from the bottom in the liquid form, the 
liquid-gas heat exchange happens in the PHE instead. Based on the measurement of cooling power with an inlet flow of 30~slpm, we determined that the corresponding efficiency is 95.3\%. 

\subsection{Gas loading and the optimization}
\label{sec:5:2}

Gas loading $G$ is a parameter to evaluate the risk of flooding~\cite{textbook,Lihulin}, defined as
\begin{equation}
\label{eq:5:2:1}
\begin{split}
G &= u_{\rm gas} \cdot \sqrt{\rho _{V}} \,,
\end{split}
\end{equation}
where $u_{\rm gas}$ is the upstream flow velocity of the xenon gas, which can be obtained by the feeding mass flow rate and the reboiler heating power.

In our test, the maximum test flow rate of 56.5~kg/h and the maximum reboiler heating power of 251~W are achieved without flooding. 
Therefore, the allowable gas loading is estimated to be at least 0.35~$\sqrt{\rm Pa}$. At this allowable gas loading, the flow rate can be increased to 60.8~kg/h with a reflux ratio of 0.1 for the radon removal, to further improve the detector radon reduction factor. For the krypton removal, the maximum feeding flow is limited by the flow controller (50~slpm) to be 17.7~kg/h. The reflux ratio can be as large as 55 with the 0.35~$\sqrt{\rm Pa}$ gas loading, under which the krypton concentration could be potentially improved to 2$\times10^{-15}$~mol/mol according to figure~\ref{fig:2:2:1:2}.

\subsection{Production krypton measurement}
\label{sec:5:3}

A purity analysis system was constructed based on a residual gas analyzer (RGA) and the liquid nitrogen cold trap~\cite{Krmeasurement}, coupled to the distillation tower. The RGA was calibrated by Kr-Xe mixtures with known Kr concentrations. Details of this system will be discussed in a separate paper~\cite{wumengmeng}. Here we only highlight results related to the commissioning of the distillation tower. For the commercial xenon used in the commissioning (krypton concentration at 0.5~ppm), we do not observe trace of krypton in the product xenon. The upper limit of the krypton concentration is set at $<$8.0~ppt at 90\% C.L.

\section{Conclusion}
\label{sec:6}

A high efficiency cryogenic distillation system is designed to remove krypton and radon for the PandaX-4T dark matter experiment. The same column can satisfy the krypton and radon removal by adjusting the running mode. By design, the krypton concentration of the product xenon is expected to be reduced by seven orders of magnitude with a 10~kg/h mass flow rate, and a radon reduction factor of about 1.8 is anticipated under a flow rate of 56.5~kg/h with the detector radon emanation considered.
The as-built tower has gone through its first commissioning run including both modes under different operation parameters. The krypton concentration of the product xenon is measured with an upper limit of 8.0~ppt, limited by the sensitivity of the purity analysis system. The radon reduction factor can be tested during the PandaX-4T operation.

\acknowledgments

The authors would like to thank the supports of the PandaX-4T collaboration. 
This project is supported by grants from the Ministry of Science and Technology of China (No. 2016YFA0400301 and 2016YFA0400302), a Double Top-class grant from Shanghai Jiao Tong University, grants from National Science Foundation of China (Nos. 11435008, 11505112, 11525522, 11775142 and 11755001), grants from the Office of Science and Technology, Shanghai Municipal Government (Nos. 11DZ2260700, 16DZ2260200, and 18JC1410200), and the support from the Key Laboratory for Particle Physics, Astrophysics and Cosmology, Ministry of Education. This project is supported by Sichuan Science and Technology Program (No.2020YFSY0057). 
We also thank the sponsorship from the Chinese Academy of Sciences Center for Excellence in Particle Physics (CCEPP), Hongwen Foundation in Hong Kong, and Tencent Foundation in China. Finally, we thank the CJPL administration and the Yalong River Hydropower Development Company Ltd. for indispensable logistical support and other help.


\end{document}